\documentclass[a4paper,12pt]{article}

\usepackage{epsfig}

\def\text#1{\mbox{#1}}
 
\newcommand{\nn}{\noindent}
\newcommand{\non}{\nonumber}
\newcommand{\tb}{\mbox{tan$\beta$}}
\newcommand{\s}{\hat{s}}

\newcommand{\ra}{\rightarrow}
\newcommand{\be}{\begin{equation}}
\newcommand{\ee}{\end{equation}}

\newcommand{\pht}{\phantom{\frac{1}{1}}\!\!\!\!}

\begin{document}

\begin{titlepage}

\begin{flushright}
ITP-Budapest 554 \\
December 1999 \\
%hep-ph ????
\end{flushright}

\vspace{1cm}

\begin{center}
\baselineskip25pt

{\large\sc Resonance Production of Three Neutral Supersymmetric \\ 
	Higgs Bosons at LHC}

\end{center}

\vspace{1cm}

\begin{center}
\baselineskip12pt

\def\thefootnote{\fnsymbol{footnote}}

{\sc G.~Cynolter, E.~Lendvai
and G.~P\'ocsik} \\ 
\vspace{1cm}

 Institute for Theoretical Physics \\
  E\"otv\"os Lorand University, Budapest

\vspace{0.3cm}
 
\end{center}

\vspace{2cm}

\begin{abstract}
  \normalsize \nn 
  Multiple production  of  Higgs particles is
  essential to study Higgs self-couplings at future
  high--energy colliders. 
  In this paper we calculated the resonance contributions to 
  the production of three lightest neutral supersymmetric
  Higgs bosons in gluon fusion at LHC. The cross sections 
  due to trilinear Higgs couplings is sizeable but the measurement 
  of the quartic coupling $\lambda_{hhhH(h)}$
  seems to be impossible.

\end{abstract}

\end{titlepage}

\def\thefootnote{\arabic{footnote}}
\setcounter{footnote}{0}

\setcounter{page}{2}

\section{Introduction}

One of the basic open questions of particle physics is the
nature of electroweak symmetry breaking. 
Beyond the discovery of the Higgs boson(s), however,
reconstructing the  Higgs potential will be necessary,
and that requires the experimental study of the
self--couplings of the Higgs bosons.

Pair production of neutral Higgs particles in gluon fusion 
sensitive to the trilinear couplings of  Higgs bosons was studied
in the SM [1] and the Minimal Supersymmetric Standard Model (MSSM) [2--5].
QCD corrections were  included in the limiting case of a 
heavy top mass \cite{2hqcd}.

In the SM  the trilinear and the quartic
couplings of the physical Higgs particle are fixed by the
Higgs mass. There is no resonance in the pair production of 
SM Higgses, and the cross section is only 
$\simeq $20--50 fb in the intermediate mass range [3].
No resonance effect is present in the production of three 
Standard Model Higgs bosons via one Higgs  in gluon fusion and it
is estimated to be under the discovery limit at LHC. 

In the MSSM there are five Higgs bosons ($h,H,A,H^\pm$)
and many trilinear and quartic couplings among them.
Two parameters describe the Higgs sector of the MSSM:
$M_A$, the mass of the CP-odd Higgs boson A, and 
 tan$\beta$, the ratio of the two
vacuum expectation values. 
For  a wide range of $M_A$, in $pp$ collision 
only the cross section  of the hh production is large [2,3].
Possible other processes, such as the gauge boson fusion
and Higgs--strahlung off W bosons or heavy quarks (top)
provide less event number.

In order to provide further tests of trilinear and quartic 
Higgs self--couplings, in this paper we calculate 
the resonance enhanced production of three lightest 
supersymmetric Higgs bosons (h) in gluon fusion
\begin{equation}
pp \to gg \to hhh .
\end{equation}

Generally four different graphs contribute to the production
of three Higgses. Gluons couple to  triangle, box or pentagon
quark loops emitting 1,2 or 3 Higgs bosons, as is seen in 
Fig.1. Squark loops are neglected.
We have omitted the pentagon graph producing a flat 
continuum background for the resonance production and
we will give the complete cross section valid in a wider range 
in a subsequent publication \cite{prep}.
We learn that the quartic coupling is not accessible by LHC experiment.
The resonance contribution to (1), however, sizeable at LHC
For instance at $ \tan \beta=3$ it yields a cross section of about 300 fb.
This is about $1/10$ times the corresponding $hh$--case [3].

The masses, widths and the couplings were calculated using the complete
one--loop and the leading two--loop radiative corrections
from \cite{carena}.
The relevant Higgs self--couplings are the following

\begin{eqnarray}
& \lambda_{hhh}& = 3\cos(2\alpha) \sin(\beta+\alpha)
+ \frac{3 \epsilon}{M_Z^2} \frac{\cos^3\alpha}{\sin \beta} \non \\
& \lambda_{Hhh}& =2 \sin(2\alpha) \sin(\beta+\alpha) 
                - \cos(2\alpha) \cos(\beta+\alpha)
+ \frac{3 \epsilon}{M_Z^2} \frac{\sin\alpha\cos^2\alpha}{\sin \beta} \non \\
& \lambda_{HHh}& =- 2 \sin(2\alpha) \cos(\beta+\alpha) 
                - \cos(2\alpha) \sin(\beta+\alpha)
+ \frac{3 \epsilon}{M_Z^2} \frac{\sin^2\alpha\cos\alpha}{\sin \beta} \non \\
%& \lambda_{HHH} = 3\cos(2\alpha) \cos(\beta+\alpha)
%+ \frac{3 \epsilon}{M_Z^2} \frac{\sin^3\alpha}{\sin \beta} \non \\
& \lambda_{Hhhh} & = 3\sin(2\alpha) \cos(2\alpha)
+ \frac{\epsilon}{8M_Z^2} \frac{\cos\alpha\sin^2\alpha}{\sin^2 \beta} \non \\
& \lambda_{hhhh} & = 3 \cos^2(2\alpha)
+ \frac{\epsilon}{4 M_Z^2} \frac{\sin^4\alpha}{\sin^2 \beta}
\label{couplings}
\end{eqnarray}

The trilinear and quartic couplings are normalized to
$\lambda_3 = [\sqrt{2} G_F]^{1/2} M_Z^2$ and
$\lambda_4 = \sqrt{2} G_F  M_Z^2$, respectively. 
The couplings depend on
$\beta$ and the mixing angle $\alpha$ of the CP--even Higgs sector
\begin{equation}
\tan 2 \alpha=
 \frac{ M_{A}^2 + M_Z^2 }{ M_{A}^2 - M_Z^2 + \epsilon / \cos 2\beta} 
 \; \tan 2 \beta .
\end{equation}
The leading $m_t^4$ one--loop corrections [9] are parametrized by
\begin{equation}
\epsilon = \frac{3 G_F}{\sqrt{2} \pi^2}
              \frac{m_t^4}{\sin^2 \beta} \;
              \log\left[ 1 + \frac{M_S^2}{m_t^2} \right].
\end{equation}
The common squark mass is fixed to $M_S$= 1 TeV.  
The  quartic couplings depend on $M_A$ and are shown for two 
values of $\tb=3$ and 30 in Fig.2.

The paper is organized as follows. In the next section 
we give the cross section of the resonance production of three h's
in gluon fusion and  in section 3 we show the numerical results.

\section[]{The cross section}

Three different channels contribute to the resonance production
of three lightest supersymmetric Higgs particles (h) in gluon fusion.
(a) One virtual Higgs boson (h,H) is produced by the heavy quark
triangle and it decays into 3 h's via the quartic coupling
$\lambda_{Hhhh}$ or $\lambda_{hhhh}$, Fig. 1a. 
(b) One virtual Higgs boson decays into 3 h's in two steps testing
trilinear couplings, $gg \rightarrow H,h \ra (H,h) h \ra hhh$, Fig. 2b.
(c) Heavy quark box diagram couples to two Higgses, one of them
decays subsequently into two h's, $gg \ra (H,h) h \ra hhh$, Fig. 2c.
We have omitted the pentagon graphs contributing only  to the 
continuum production. 
We believe  it increases the production at large $\tb$
due to the large hbb--coupling, but does not
change our conclusion about the measureability of the
quartic Higgs--couplings.

There are two different helicity amplitudes contributing to the
total cross section: the  total spin of the two gluons along
the collision axis can be $S_z=0$ (contribution F) or
 $S_z=2$ (contribution G).
The triangle graphs give only  contributions $F_3$ ($ S_z=0$) as they
involve a single spin--0 Higgs intermediate state.
Box graphs give both contributions $F_4$ ($ S_z=0$) and  $G_4$ ($S_z=2$).
$F_4$ and $G_4$ are Lorentz and gauge invariant decompositions
of the box amplitude. The tensor basis is the following

\begin{eqnarray}
& &S_z=0 \quad A^{\mu\nu} = g^{\mu\nu}-\frac{p_1^\nu p_2^\mu }{(p_1 p_2)}  \\
& & S_z=2 \quad B^{\mu \nu} = g^{\mu \nu}
               +\frac{1}{p_T^2 (p_1 p_2)} \bigl ( p_3^2 p_1^\nu p_2^\mu
               -{2 (p_2 p_3) p_1^\nu p_3^\mu} -{2 (p_1 p_3) p_2^\mu p_3^\nu}
               +{2 p_3^{\mu} p_3^{\nu} (p_1 p2)}  			
	 \bigr ) \non \label{tens}  
\end{eqnarray}
where $p_1,p_2$ are the momenta of the incoming gluons and
$p_3$ is one of the momenta of the outgoing Higgs bosons
($p_3,p_4,p_5$). Here
$ p_T^2 =2\frac{(p_1p_3)(p_2p_3)}{(p_1p_2)}-p_3^2 $ 
is the transverse momentum of the third particle.
%$\epsilon(p_1,p2,p3,\mu)= \epsilon_{\mu \rho \sigma \kappa} p_1^{\rho} p_2^{\sigma} p_3^{\kappa}$. 
Tensors $ A^{\mu\nu}$ and $ B^{\mu \nu}$  are orthogonal and normalized to $2$ [2].

The M--matrix of the process is
\be
{\cal M} =
        \frac{(\sqrt{2}G_F)^{3/2} \alpha_s \s }{4  \pi} \;
        \epsilon_a^\mu \epsilon_b^\nu \; \delta_{ab}         
         ( F A_{\mu \nu} + G B_{\mu \nu} ).
\ee
The spin and color averaged parton level cross section of the process $gg \ra hhh$ is
% Here all the factors from the coupling c's are multiplied!!!!
\begin{equation}  
{d\hat\sigma (gg\to hhh)} = 
\frac{\sqrt{2}G_F^3\alpha_s^2} {1024 (2\pi)^6} 
    \Big[ | F |^2 + |G|^2 \Big]	
  \frac{\lambda^{1/2}(\s_{45},p_4^2,p_5^2)}{\s_{45}} 
	d\s_{45}d\s_{13} d\Omega^{CM}_{45}.
\label{sigma}
\end{equation} 

Here we used the Chew-Low parametrization of the three particle 
phase space \cite{kajan}, a trivial angle integration was carried out,
$\s_{ik}=(p_i+p_k)^2$,
and the usual lambda function is 
$\lambda(x,y,z)=\left (x-(\sqrt{y}+\sqrt{z})^2 )
	(x-(\sqrt{y}-\sqrt{z})^2 \right) $.
$ d\Omega^{CM}_{45} $ is the differential solid angle in the center 
of mass system of particles 4 and 5.

$F$ receives contributions from the triangle and box graphs,
while only box diagrams give contributions $S_z=2$  to $G$
\begin{eqnarray}
	F &  =&\sum_{t,b}(  C_3 \; F_3 \; + 
	\sum_{i=3,4,5}C^{(i)}_4\; F^{(i)}_4)  \non \\
	G & = & \sum_{t,b}\sum_{i=3,4,5} C^{(i)}_4 G^{(i)}_4 .
\label{fsum}
\end{eqnarray}
Here we separated the functions $F_3,F_4,G_4$ responsible 
for the heavy quark triangle and boxes, $C_3,C_4$ 
contain the coupling constants and the propagators. 
The amplitude has to be summed over all quark
flavours. 
However, the four light quarks can be neglected having very 
small Higgs couplings.
Supersymmetric particles are assumed to be too heavy to
participate in the triangle or box loop.
The second sum in the box contribution corresponds to the
outgoing Higgs particle that couples directly to the quark loop.

We calculated $F_3,F_4,G_4$ 
%with the techniqe of Ref. \cite{veltpas}
and found them in agreement with the results of \cite{spira}, \cite {bij}.
For instance,
\be
F_3 = 2 \frac{m_Q^2}{\s} \left ( 2+ (4m_Q^2-\s) C_{12} \right), \quad 
\s=\s_{12}.
\ee
where
\be
C_{ij} =  \int \frac{d^4q}{i\pi^2}~\frac{1}
{(q^2-m_Q^2)\left[\pht (q+p_i)^2-m_Q^2\right]
\left[\pht (q+p_i+p_j)^2-m_Q^2\right]}.
\ee

$F_4, G_4$ have long analytic expressions 
which can be found in ref \cite{spira}. 
The box functions $F^{(i)}_4, G_4^{(i)}$ in (\ref{fsum}) 
depend on three momenta 
$(p_1,p_2, p_i),\quad i=3,4,5 $ and $F_4^{(i)}=F_4(p_1,p_2,p_i)$ etc.

In case of a large quark mass ($m^2_Q \gg \s \sim M^2_{h,H}$)
we get as in [2]
\be
F_3=\frac{2}{3}+{\cal O}\left(\hat s/m_Q^2\right), \qquad
 F_4^{(i)} =-\frac{2}{3} \quad {\cal O}\left( {\s}/m_Q^2\right),
  \quad G_4^{(i)} ={\cal O}\left( {\hat s}/m_Q^2\right)
\ee
In the limit of light quark masses ($m_Q^2 \ll \s \sim M_{h,H}^2$)
all the form factors vanish  to ${\cal O}\left(m_Q^2/\s\right)$.

The generalized triangle couplings are the following
\begin{eqnarray}     
&&  C_3 = C_3^h + C_3^H+ C_3^{hh} + C_3^{hH}+ C_3^{Hh} + C_3^{HH} \non \\
&&  C^{H_a}_3 = \lambda_{hhhH_a} \frac{M_Z^2}
	{\s-M_{H_a}^2+iM_{H_a}\Gamma_{H_a}} g_Q^{H_a}  \\
&&  C^{H_aH_b}= \lambda_{H_aH_b h} \frac{M_Z^2}
   {\s-M_{H_a}^2+iM_{H_a}\Gamma_{H_a}}   \lambda_{H_b h h} \frac{M_Z^2}
   {\s-M_{H_b}^2+iM_{H_b}\Gamma_{H_b}} g_Q^{H_a}, \quad H_{a,b}=h/H. \non
\end{eqnarray}
The box couplings are
\begin{eqnarray}
 && C_4 = C_4^h + C_4^H   \non \\
 &&  C^{(h,H)(i)}_4 = \lambda_{hh (h/H)} \frac{M_Z^2}
	{\s_{kl}-M_{h/H}^2+iM_{h/H}\Gamma_{h/H}} g_Q^h  g_Q^{h/H},
 \end{eqnarray}
$i,k,l$ are cyclic. $g_Q^{h/H}$ denotes the Higgs--quark couplings 
normalized to the SM Yukawa coupling $[\sqrt{2}G_F]^{1/2}m_Q$,
\be
g_t^h=\cos\alpha/\sin\beta, \quad g_b^h =-\sin\alpha/\cos\beta,
\quad
g_t^H=\sin\alpha/\sin\beta,  \quad g_b^H=\cos\alpha/\cos\beta.
\ee

\section{Results}

 \indent Now, we calculate the production cross section of three lightest 
neutral  supersymmetric Higgs bosons 
$pp \ra hhh + X$ via gluon fusion at the
LHC energy of $\sqrt{s}=14$ TeV in the above picture
by folding the parton level fusion cross section with the
gluon luminosity.
We used the GRV structure functions \cite{pdgrw} for the gluon luminosity
at $Q^2=\s$. The numerical integration was made by the VEGAS package 
\cite{vegas}. The total cross section vs. $M_A$ 
is shown in Fig.3 for two represantative
values of tan$\beta =3$ and $\tb= 30$.
In the plotted range of $M_A$ the heavy CP--even
 Higgs boson (H) is nearly degenerate in mass with A while the mass of the
lightest neutral Higgs h is quickly approaches approximately 104
 GeV from below.

For  $\tb\!=\!3$ in the range of $M_A =$ 200--350 GeV 
 the cross section is enhanced by resonance effect and large.
The cross section goes up to nearly 300 fb and slowly decreases
to about 150 fb giving a large number of events 
at the expected luminosity $\int L dt \simeq 100 fb^{-1}/year$.
The main contribution comes from the box diagram where the heavy CP--even
 Higgs boson H couples to the quark loop and decays 
into two bosons h. $M_h$ slowly increases around 100 GeV when
$M_H \simeq M_A$ reaches 200 GeV the H propagator enters 
 the resonance  region. 
At $M_H \sim 2 m_t$ there is a threshold effect due to the fall--off of the
branching ratio $BR(H \ra hh)$  and partly because on--shell
top quark pairs can be produced in the quark loop.
For large values of  $M_A$ the cross section reaches a continuum value of
a few fb. The dip at small $M_A$, similarly  to the case of $\tb =30$, 
is the result of the zero  in the trilinear couplings $\lambda_{(H/h)hh}$.

For $\tb\!=\!30$ the main contribution comes from the box diagram where
an h couples to the quark loop, propagates 
and decays into two h's via $\lambda_{hhh}$. 
The Hhh coupling is small
compared to the case of $\tb =3$ while the hhh coupling is 
larger giving a sizeable continuum. 
There is no observable resonance effect and the cross section is
nearly constant, 6 fb, versus $ M_A $ after the coupling constant changed sign.
The K factor is expected to increase the cross section as in other cases [6,3].

The contributions of the diagrams containing 
a heavy quark triangle are also shown
in Fig.3 at $\tb\!=\!3$. 
The lowest  curve corresponds to the diagram with quartic couplings,
 the one in the middle sums up all the contributions of the graphs involving
a quark triangle. There is a clear resonance effect in both curves  when
$M_H$ reaches $3 M_h$ at $\sim$310 GeV. Here the first propagator of
$ gg \ra H  \ra H h\ra hhh$ becomes resonant too. The first rise in
the middle curve is the result of the resonance in the second propagator
of Fig. 2b, similarly to the box.
The triangle loops yield, however, only a small fraction of the cross section
even at their peak at 300-350 GeV.

The suppressed contribution of the quartic coupling implies that
in $pp \ra hhh+ X$ the measurement of the quartic coupling is
 not possible. Not only the box contribution is larger
by two orders of magnitude but also the other triangle diagrams 
are 10 times larger. For $\tb=30$ the case is even worse.

In conclusion, in this paper we have calculated the resonance 
contribution to the cross section of 3h--production at LHC via gluon fusion. 
The main contribution comes from the trilinear 
Higgs coupling while the quartic ones turn out to be negligible. 
In the resonance region at small $\tb$ the 
ratio of 2h and 3h production is about 10.

This work was partially supported by 
Hungarian Science Foundation Grants under Contract
Nos. OTKA T029803 and F022998.

%\vspace*{2cm}
\newpage

\nn
{\large \bf FIGURE CAPTIONS}
\\

\nn {\bf Fig.1:} Generic diagrams contributing to the production of
three CP--even MSSM Higgs bosons in gluon--gluon collisions, $gg\to hhh$:
 a) triangle  with quartic coupling,  b)triangle with trilinear couplings
c) box , d) pentagon contributions.
\\

\nn {\bf Fig.2:} Quartic Higgs couplings in the MSSM as  functions
of the pseudoscalar Higgs mass $M_A$ for two representative values of 
$\tb\!=\!3$ and 30.\\

\nn {\bf Fig.3:} Total cross sections for production of three lightest
CP--even MSSM Higgs bosons in gluon--gluon collisions at the LHC
for  $\tb=3$ and 30 (upper two curves). 
The upper axis presents the scalar Higgs masses $M_{h}$ for  $\tb=3$
corresponding to the pseudoscalar masses $M_A$. 
The lower two dashed curves represent the resonance contributions of
the triangle graphs of Fig. 2a and 2b for $\tb\!=\!3$.
\\

\newpage

%%%%%%%%%%%%%%%%%%%%%%%%%%%%%%%%%%%%%%%%%%%%%%%%%%%%%%%%%%%%%%
\begin{figure}[ht!]
\begin{center}
 \mbox{
\psfig{figure=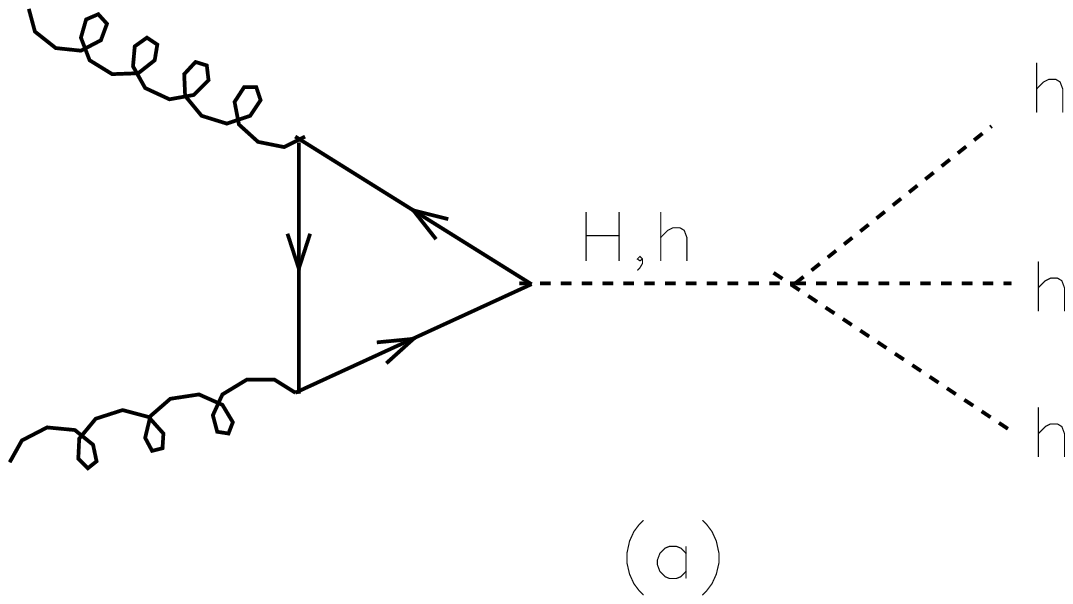,width=7cm,height=7cm}}
\hspace{1em}
\mbox{
\psfig{figure=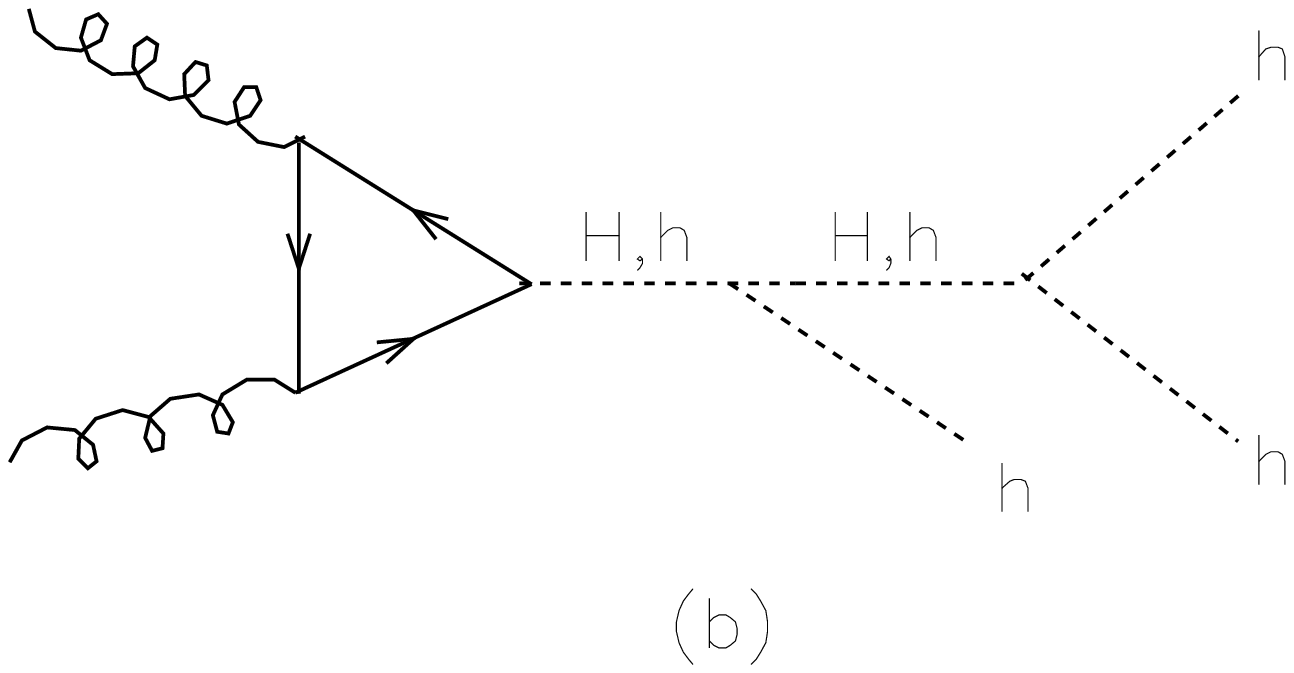,width=7cm,height=7cm}}
\end{center}
\vspace{-1.5em}
%\caption[]{Triangle diagrams with quartic coupling }
\begin{center}
 \mbox{
\psfig{figure=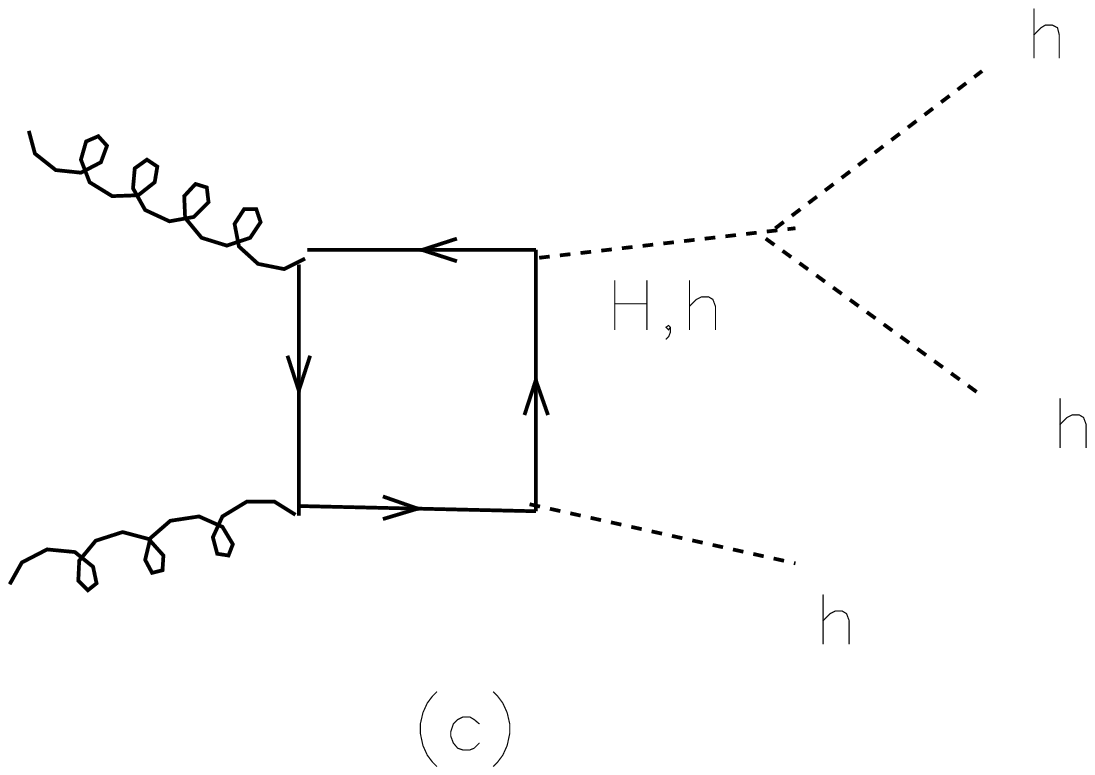,width=7cm,height=7cm}
%\hspace{1em}
%\end{center}
\hspace{1em}
%\begin{center}
% \mbox{
\psfig{figure=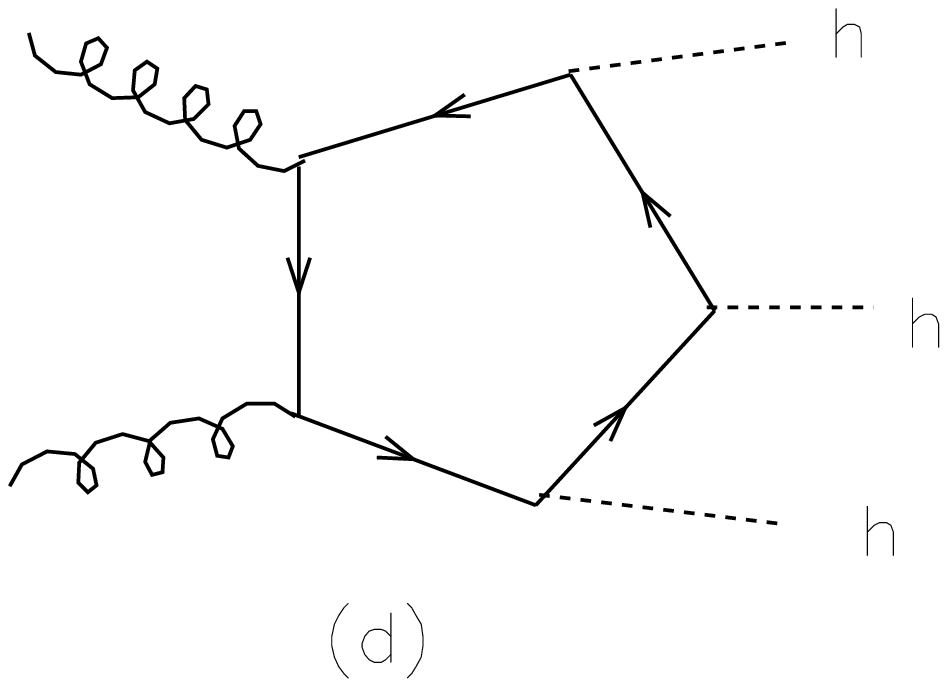,width=7cm,height=7cm}}
\hspace{1em}
%\caption[]{}
\mbox{Fig. 1}
\end{center}

\label{fig1}
\end{figure}
%%%%%%%%%%%%%%%%%%%%%%%%%%%%%%%%%%%%%%%%%%%%%%%%%%%%%%%%%%%%%%

\newpage

%%%%%%%%%%%%%%%%%%%%%%%%%%%%%%%%%%%%%%%%%%%%%%%%%%%%%%%%%%%%%%
\begin{figure}[ht!]
\begin{center}
 \mbox{
\psfig{figure=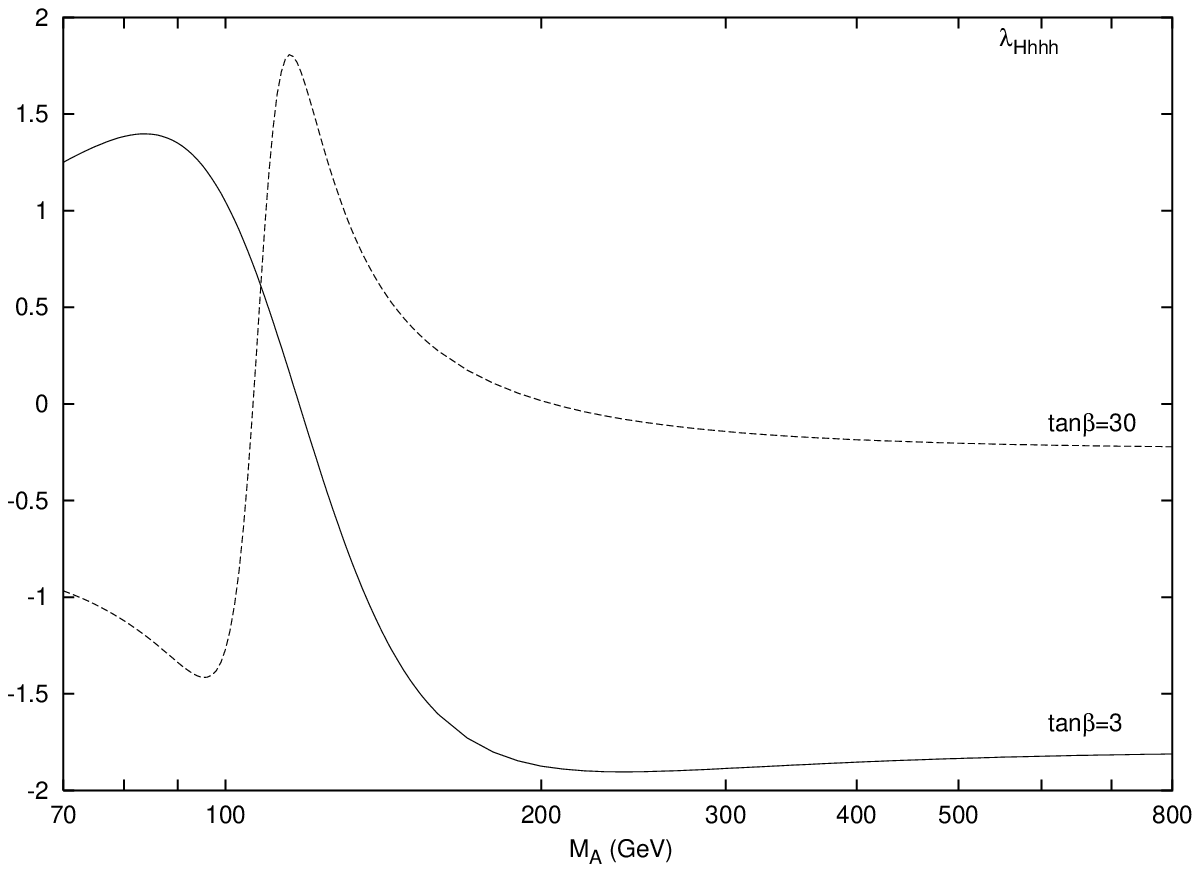,width=15cm,height=10cm}}
%\hspace{1em}

\mbox{Fig. 2a}
\end{center}
%\vspace{1em}

\begin{center}
 \mbox{
\psfig{figure=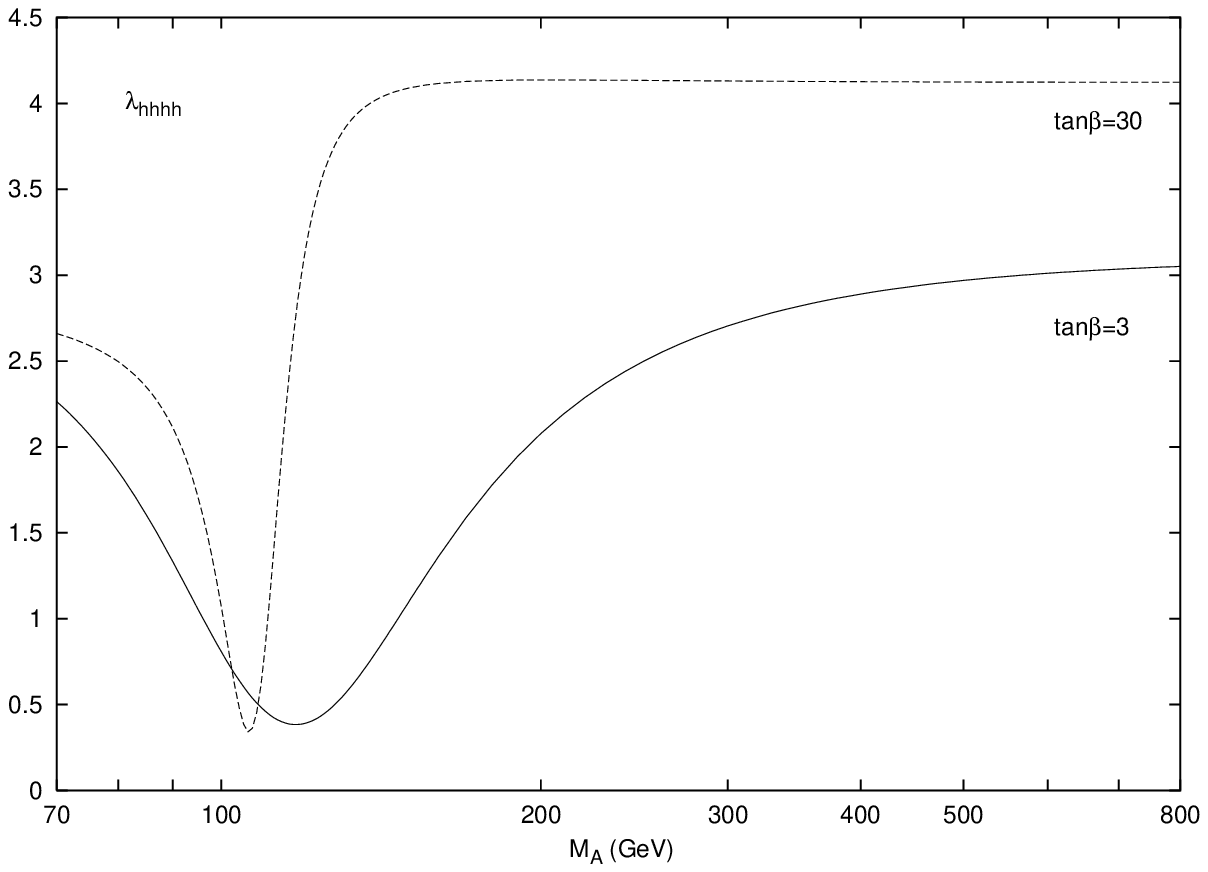,width=15cm,height=10cm}}

%\caption[]{}
\mbox{Fig. 2b}
\end{center}

\label{fig2}
\end{figure}
%%%%%%%%%%%%%%%%%%%%%%%%%%%%%%%%%%%%%%%%%%%%%%%%%%%%%%%%%%%%%%

\newpage

\vspace{5cm}
%%%%%%%%%%%%%%%%%%%%%%%%%%%%%%%%%%%%%%%%%%%%%%%%%%%%%%%%%%%%%%
\begin{figure}[ht!]
\begin{center}
 \mbox{
\psfig{figure=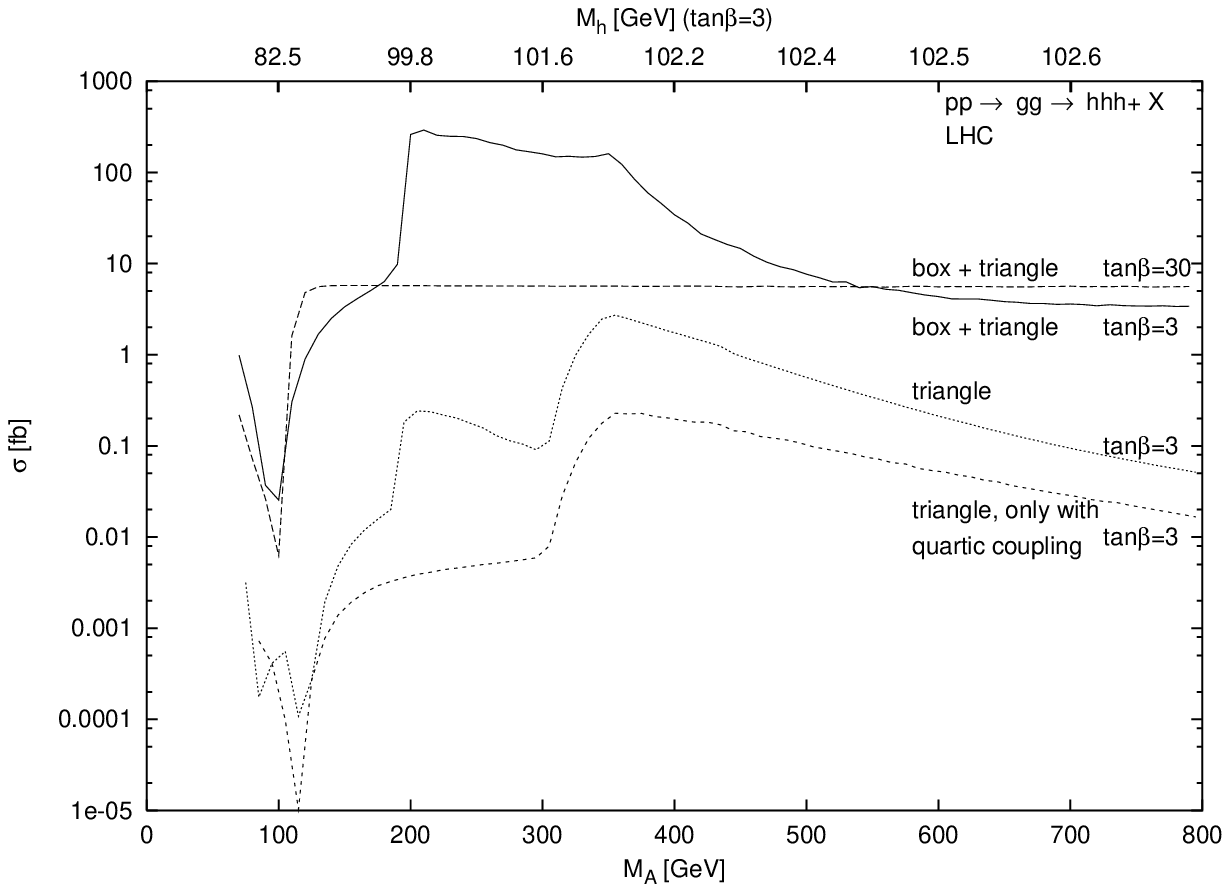,width=16cm,height=13cm}}
%\hspace{1em}

\mbox{Fig. 3}
\end{center}

\label{fig3}
\end{figure}
%%%%%%%%%%%%%%%%%%%%%%%%%%%%%%%%%%%%%%%%%%%%%%%%%%%%%%%%%%%%%%

\end{document}